\documentclass[preprint,12pt]{elsarticle}

\newtheorem{thm}{Theorem}
\newtheorem{lem}[thm]{Lemma}
\newdefinition{rmk}{Remark}
\newproof{pf}{Proof}
\newproof{pot}{Proof of Theorem \ref{thm2}}
\usepackage{amssymb}

\journal{Elsevier}

\begin{document}

\begin{frontmatter}



\title{Existence and uniqueness of generalized monopoles in six-dimensional non-Abelian gauge theory}


\author[MKC]{S.X.Chen\corref{cor1}\fnref{fn1}}
\ead{chensx1982@gmail.com}

\author[AKC]{L.Pei}
\ead{peilongsf@gmail.com}
\cortext[cor1]{Corresponding author}

\fntext[fn1]{The research of the first author was supported in part
by the Natural Science Fund of Henan Education Office (2007110004)
and (2008A110002).}

\address[MKC]{School of Mathematics and Information Sciences, Henan University, Minglun Street, Kaifeng, P.R.China 475001}

\address[AKC]{Arts and Science Experimental Class, Henan University, Minglun Street, Kaifeng, P.R.China 475001}

\begin{abstract}
In this paper, we established the existence and uniqueness of the
spherically symmetric monopole solutions in ~$SO(5)$~  gauge theory
with Higgs scalar fields in the vector representation in
six-dimensional Minkowski space-time and obtain sharp asymptotic
estimates for the solutions. Our method is based on a dynamical
shooting approach that depends on two shooting parameters which
provides an effective framework for constructing the generalized
monopoles in six-dimensional Minkowski space-time.
\end{abstract}


\end{frontmatter}


\section{Introduction}
Long time ago, Dirac showed that quantum mechanics admits a magnetic
monopole of quantized magnetic charge despite the presence of a
singular Dirac string\cite{Dirac,WY2}. Much later, G. 't Hooft and
Polyakov showed that such magnetic monopoles emerge as regular
configurations in $SO(3)$ gauge theory with spontaneous symmetry
breaking triggered by triplet Higgs scalar
fields\cite{GH,AMP,GT,PS}. Although a monopole has not been detected
or produced experimentally as a single particle, the existence of
such objects has far reaching consequences.In the early universe,
monopoles might have beed copiously produced, which vanished due to
various physical interaction such as Pair production caused by
Coulomb interaction among monopoles and anti-monopoles\cite{DS} but
have significantly affected the history of the universe since then.
For example, monopoles magnetic monopoles flow dispute the dynamo
action, leading to a slow dynamo action in the best hypothesis or a
decay of the magnetic field\cite{AG}. Also, Monopoles played a very
important role in the formation of galaxy formation\cite{JeM}. G¡¯t
Hooft-Polyakov monopoles emerge in grand unified theory of
electromagnetic, weak, and strong interactions as well.

It is important and promising to explore solitonic objects in higher
dimensional space-time under the super-string scenario, and recent
extensive study of domain walls in super-symmetric theories, for
instance, may have a direct link to the brane world
scenario\cite{YMK1,YMK2,YMK3}. The energy of 't Hooft-Polyakov
monopoles is bound from below by a topological charge. Monopole
solutions saturate such bound, thereby the stability of the
solutions being guaranteed by topology\cite{Bogo}. This observation
prompts a question that if there can be a monopole solution in
higher dimensions. Kalb and Ramond introduced Abelian tensor gauge
fields coupled to closed strings\cite{KR}. Nepomechie showed that a
new type of monopole solutions appear in those Kalb-Ramond
antisymmetric tensor gauge fields\cite{RIN}. Their implications to
the confinement\cite{RSav,PO,RBP} and to ten-dimensional Weyl
invariant space-time \cite{YH} has been explored. Topological
defects in six dimensional Minkowski space-time as generalization of
Dirac's monopoles were also found\cite{CNY}. Tchrakian has
investigated monopoles in non-Abelian gauge theory in higher
dimensions whose action involves polynomials of field strengths of
high degrees\cite{DHT,DHTZ}. Furthermore, it has been known that
magnetic monopoles appear in the matrix model in the gauge
connections describing Berry's phases on fermi states. In
particular, in the $USp$ matrix model they are described by
~$SU(2)$~-valued anti-self-dual connections\cite{IM,ChenIK}.
H.Kihara and his team presented regular monopole configurations with
saturated Bogomolny bound in $SO(5)$ gauge theory in six
dimensions\cite{KHN}. Self-gravitating Yang monopoles in all
dimensions was also studied by G.W.Gibbons and
P.K.Townsend\cite{GT}.The purpose of this paper is to establish an
existence and uniqueness theorem for these generalized monopoles in
six-dimensional non-Abelian gauge.

In the next section, we first briefly discussed the mathematical
structure of the problem of the existence of generalized monopoles
in six-dimension non-Abelian gauge. We then state our main existence
and uniqueness theorem for these solutions. In the third section, we
transform the first-order equations into a second-order non-linear
equation, and then we introduce a series of variable transformations
to reduce the equation into a linear equation. In this case, the
existence of generalized monopole solutions is seen to be equivalent
to the existence problem of a nonlinear two-point boundary value
problem. In section 4, we present a dynamical shooting method which
solves the existence problem completely and may be used as a
constructive method for numerical computation. We shall also obtain
sharp asymptotic estimates for the solutions.
\section{Mathematical Structure and Theorem}
Following  Kihara, Hosotani and Nitta, we recall that a key to find
correct Bogomolny equations in six-dimensional space-time is
facilitated with the use of the Dirac or Clifford algebra. Let's
consider ~$SO(5)$~ gauge theory in six-dimensional space-time. Based
on Clifford algebra and gauge transformation, the action is given by
$$
\begin{array}{rcl}
I&=&
\int[-\frac{1}{8\cdot{4!}}TrF^{2}*F^{2}-\frac{1}{8}TrD_{A}\phi-\frac{\lambda}{4!}(\phi^{a}\phi_{a}-H_{0}^{2})^2d^{6}x]\\
&=&\int
d^{6}x[-\frac{1}{8\cdot{4!}}Tr(F^{2})_{\mu\nu\rho\sigma}(F^{2})^{\mu\nu\rho\sigma}-\frac{1}{2}D_\mu\phi^{a}D^{\mu}\phi_{a}+\lambda(\phi^{a}\phi_{a}-H_{0}^{2})^{2}],
\end{array}
\eqno(2.1)
$$
where the components of $
F^{2}=\frac{1}{8}\{F_{\mu\nu},F_{\rho\sigma}\}dx^{\mu}\land dx^{\nu}
\land dx^{\rho} \land dx^{\sigma} $ are given by
$$
\begin{array}{lr}
(F^{2})_{\mu\nu\rho\sigma}~=~T_{\mu\nu\rho\sigma}^{e}\gamma_{e}-S_{\mu\nu\rho\sigma},\\
T_{\mu\nu\rho\sigma}^{e}(A)~=~\frac{1}{2\cdot4!}\epsilon^{abcde}(F_{\mu\nu}^{ab}F_{\rho\sigma}^{cd}+F_{\mu\rho}^{ab}F_{\sigma\nu}^{cd}+F_{\mu\sigma}^{ab}F_{\nu\rho}^{cd}),\\
S_{\mu\nu\rho\sigma}(A)~=~\frac{1}{4!}(F_{\mu\nu}^{ab}F_{\rho\sigma}^{cd}+F_{\mu\rho}^{ab}F_{\sigma\nu}^{cd}+F_{\mu\sigma}^{ab}F_{\nu\rho}^{cd}).
\end{array}
\eqno(2.2)
$$
The canonical conjugate momentum fields are given by
$$
\begin{array}{rcl}
\Pi_{i}^{ab}&=&\frac{\delta I}{\delta A_{i}^{ab}}\\
&=& \frac{1}{3!}T_{0jkl}^{e}\frac{\delta
T_{0jkl}^{e}}{\delta F_{0i}^{ab}}+\frac{4}{3!}S_{0jk}\frac{\delta S_{0jkl}^{e}}{\delta F_{0i}^{ab}}~\\
&=& \frac{1}{3}(M_{i,jkl}^{ab,e}M_{m,jkl}^{cd,e}+N_{i,jkl}^{ab,e}N_{m,jkl}^{cd,e})F_{0m}^{cd}\\
&:=&U_{i,m}^{ab,cd}F_{0m}^{cd}~~,
\end{array}
 \eqno(2.3)
$$
where $U$ is a symmetric, positive-definite matrix. To confirm the
positivity of the Hamiltonian, we take the $A_{0}=0$ gauge in which
$F_{0i}^{ab}=A_{i}^{ab}$. It immediately follows that
$$
E=\int d^{5} x[\frac{1}{2}\Pi
U^{-1}\Pi+\frac{1}{2\cdot4!}{(T_{ijkl}^{e})^{2}+(S_{ijkl})^{2}+H_{\phi}}]\geq0~,
\eqno(2.4)
$$
where $H_{\phi}$ is the scalar field part of the Hamiltonian
density.

In the $A_{0}=0$ gauge, the energy becomes lowest for static
configurations  $A_{i}^{ab}=\phi_{a}=0$ and it is given by
$$
\begin{array}{rcl}
E&=&\int d^{5}x
\frac{1}{4!}[\frac{1}{2}(T_{ijkl}^{e} \mp\epsilon^{ijklm} D_{m}\phi^{e})^{2}+\frac{1}{2}(T_{ijkl}^{e})^{2}\\
&\pm&\epsilon^{ijklm}T_{ijkl}^{e}D_{m}\phi_{e}+\lambda(\phi_{a}\phi^{a}-H_{0}^{2})^{2}]
\\
&\geq&\pm\int d^{5} x
\frac{1}{4!}\epsilon^{ijklm}T_{ijkl}^{e}D_{m}\phi^{e}\\
&=&\pm\int TrD_{A}\phi F^{2}
\\
&:=& \frac{16\pi^{2}}{g^{2}}H_{0}\Psi .
\end{array}
 \eqno(2.5)
$$
As $D_{A}F = 0$ and thereby $TrD_{A}\phi F^{2} =d(Tr\phi F^{2})$ ,
$\Psi$ can be expressed as a surface integral
$$
\Psi=\pm\frac{g^{2}}{16\pi^{2}}\int_{S^{4}}Tr\phi F_{2}~, \eqno(2.6)
$$
where $S^{4}$ is a space infinity of $R^{5}$.

The Bogomolny bound equation is
$$
\begin{array}{l}
*_{5}(F\wedge F)=\pm D_{A}\phi~,
\end{array}
 \eqno(2.7)
$$
where $*_{5}$ is Hodge dual in five-dimensional space. In
components, it is given by
$$
\begin{array}{rcl}
\epsilon_{ijklm}T_{ijkl}^{e}&=&\pm D_{m}\phi_{e}~,\\
S_{ijkl}&=&0~.
\end{array}
 \eqno(2.8)
$$

Let us define $e ¡Ô~:=~x^{a}\gamma_{a}/r$ and make a hedgehog
ansatz[21]
$$
\begin{array}{rcl}
\phi&=&H_{0}U(r)e~,\\
A&=&\frac{1-k(r)}{2g}ed e~.
\end{array}
 \eqno(2.9)
$$
It follows immediately that
$$
\begin{array}{rcl}
D_{A}\phi&=&H_{0}(KUd e+U' e d r)~,\\
F&=&\frac{1-K^{2}}{4g}d e \wedge d e-\frac{K'}{2g}e d r\wedge de~.
\end{array}
 \eqno(2.10)
$$
Accordingly, the boundary condition is
$$
\begin{array}{l}
U(\infty)=\pm 1,~ U(0)=0,~K(\infty)=0,~K(0)=1.
\end{array}
 \eqno(2.11)
$$
Applying $*_{5}(de \wedge de\wedge de\wedge de)=\frac{4! ed
r}{r^{4}}$ and $*_{5}(ed r\wedge dr\wedge dr\wedge dr)=\frac{3!ed
r}{r^{4}}$, the Bogomolny boundary equation(2.7)(with a plus
sign)becomes
$$
\begin{array}{lr}
KU~=~-\frac{\displaystyle(1-K^2)d K}{\displaystyle\tau^2d\tau}~,\\
\frac{\displaystyle d U}{\displaystyle d\tau}~=~\frac{\displaystyle(1-K^2)^2}{\displaystyle\tau^4}~,\\
U(\infty)~=~1,~U(0)~=~0,~K(\infty)~=~0,~K(0)~=~1~,
\end{array}
\eqno(2.12)
$$
where $\tau=ar$, $a=(\frac{2g^{2}}{3} H_{0})^{\frac{1}{3}}$.\\
In this case, $U$ increases as $\tau$ so that  $U(\infty) = 1$. A
solution in the case $-D_{A}\phi=*_{5}(F\wedge F)$ is obtained by
replacing ~$U$~by~$-U$.

Our main existence and uniqueness theorem for generalized monopole
solutions in the six-dimension non-Abelian gauge theory can be
stated as follows:

\begin{thm}
For any real number $g>0$ and $H_{0}>0$, the two point boundary
value problem (2.12) has a unique solution $(K(r),U(r))$ so that
$K(r)$ is strictly decreasing and $U(r)$ is strictly increasing for
any $r>0$. Besides, there hold the sharp asymptotic estimates
$$
\begin{array}{lr}
K~=~O(e^{-Cr^{3}}),~U~=~1+O(r^{-3}),~r \to
\infty, C>0,\\
K~=~1+O(r^{2}),~U~=~O(r),~r\to 0.
\end{array}
$$

This solution uniquely gives rise to a spherically symmetric
finite-energy monopole solution of unit topological charge for
non-Abelian gauge theory in six-dimensional Minkowski space-time.

\end{thm}

\section{Second-Order Governing Equation}
Noting that the two equations in (2.12) can be combined to yield
$$
\frac{d(\frac{1-K^{2}}{\tau^{2}K})}{d \tau}
\frac{dK}{d\tau}~+~\frac{1-K^{2}}{\tau^{4}}=0, \eqno(3.1)
$$
or equivalently, in terms of $s=\ln\tau$ and $f(s)=K^{2}$:
$$
f''-\{3+\frac{f'}{f(1-f)}\}f'+2f(1-f)=0. \eqno(3.2)
$$
Accordingly, the boundary condition becomes
$$
\begin{array}{lr}
f(-\infty)=1,f(\infty)=0.
\end{array}
 \eqno(3.3)
$$

We will prove that $0<f(s)<1$, $\forall s \in (-\infty,\infty)$.
Note that $f=0$ and $f=1$ are two equilibrium solutions of equation
(3.2), thus the existence and uniqueness theorem for solutions of
ordinary differential equation allow us to see that $0<f<1$,
$\forall s\in(-\infty,\infty)$. To make it convenient for us to
solve our problem, we apply the transformation:~$G(s)=\ln f(s)$.
Under this transformation, ~$f'$~and~ $f''$~ can be represented as
follows:
$$
f'=e^{G}G,~~f''=e^{G}(G')^{2}+e^{G}G''.
 \eqno(3.4)
$$
Inserting (3.4) into (3.2), we have
$$
G''+(G')^{2}-\{3+\frac{G'}{1-e^{G}}\}G'+2(1-e^{G})=0 .
 \eqno(3.5)
$$
Meanwhile,~it is easy to see that~$-\infty<G<0$, and the boundary
condition naturally becomes as follows:
$$
G(-\infty)=0,~~G(\infty)=-\infty.
 \eqno(3.6)
$$
Furthermore, we can see that equation (3.5) can be simplified to
$$
(G-e^{G})'-3(G-e^{G})+2(1-e^{G})^{2}=0.
 \eqno(3.7)
$$
To further simplify our problem, we introduce the transformation
$V=G-e^{G}$. Since the function $V(G)=G-e^{G}$ is strictly
increasing in the interval $-\infty<G<0$, it is invertible, and its
inverse function $Q(V)$(say)~enjoys the same properties over the
interval $(-\infty,-1)$. In terms of the variable $V$, the equation
(3.7)  and its associated boundary condition becomes
$$
\begin{array}{lr}
V''-3V'$=$-2(1-e^{G})^{2},~s\in(-\infty,\infty),\\
V(-\infty)$=$-1,~V(\infty)=-\infty.
\end{array}
 \eqno(3.8)
$$

The equation (3.8) seems more tractable than equation (3.1) except
that the function $Q(V)$ is not defined for $V\geq-1$, which makes
it inconvenient to conduct a discussion. In order to fix this
problem, we will make a suitable extension of the function
$(1-e^{Q(V)})^{2}$ to $V\geq-1$. Note that
$$
\lim\limits_{V\to -1}(1-e^{Q(V)})^{2}=\lim\limits_{G\to
0}(1-e^{G})^{2}=0.
 \eqno(3.9)
$$
Moreover, for $V<-1$, we have
$$
\frac{d}{d V}(1-e^{Q(V)})^{2}=-2e^{Q(V)},
 \eqno(3.10)
$$
which tends to $-2$ as $V\to -1$.  Hence, we can modify (3.8) into
the following form,

$$
\begin{array}{lr}
V''-3V'=R(V)~:=~ \left\{\begin{array}{ll}
-2(1-e^{Q(V)})^2, & \textrm{$V<-1$},\\
4(V+1), & \textrm{$V\geq-1$}.
\end{array}
\right.
\end{array}
\eqno(3.11)
$$
We see that $R(V)$ is a differentiable function for all $V$. We will
consider~(3.11) subject to boundary condition in (3.8). Although
(3.11) alters the original equation in(3.8) due to its modified
right-hand side function, we shall obtain a solution $V(s)$ that
says negative for all $s\in (-\infty,\infty)$. In this way, we
recover a solution to the original boundary value problem (3.8) as
expected. Hence, our boundary value problem consisting of (3.11) and
boundary condition in (3.8) becomes
$$
V''-3V'=R(V),~s\in (-\infty,
\infty),~V(-\infty)=-1,~V(\infty)=-\infty,
 \eqno(3.12)
$$
where and in the sequel, we still use the prime $'$ to denote the
differentiation with respect to the variable $s$ when there is no
risk of confusion.
\section{Mathematical Analysis}
To solve the two-point boundary value problem (3.12), we use a
dynamical shooting method. This method was once used to solve
problems\cite{ChenZC,WY1} in the field of mathematical physics. When
we do this, we need to consider the initial value problem
$$
V''-3V'=R(V),~s\in (-\infty, \infty),~V(0)=m,~V'(0)=-n.
 \eqno(4.1)
$$
Since we are looking for a solution $V<-1$, we naturally assume
$$
m<-1.
 \eqno(4.2)
$$
Under the assumption (4.2), we shall show that when $n$ is suitably
chosen in (4.1), we may obtain a solution to (3.12). It can be seen
from the structure of the problem that the boundary condition
~$V(-\infty)=-1$~is a crucial part to realize. So we shall look at
this end first. For this purpose, we set $t=-s$ in the half interval
$-\infty <s\leq 0$~and convert (4.1) into the form
$$
V''+3V'=R(V),~t>0,~V(0)=m,~V'(0)=n,
 \eqno(4.3)
$$
where the prime  $'$ denotes the differentiation with respect to the
reversed variable $t$. We also use $V_{t}$ to denote $\frac{d V}{d
t}$. For fixed $m$ satisfying (4.2), we use $V(t;n)$ to denote the
unique solution of (4.3) which is defined in its interval of
existence.

We are now ready to launch a shooting analysis for (4.3). We express
the set of real numbers $R$  as the disjoint union of three data
sets as follows:

$\beta^{-}~=~\{n\in{R}|$ there exists $t>0$ so that $V_t(t;n)<0\},$

$\beta^{0}~=~\{n\in{R}|~ V_t(t;n)>0$ and $V(t;n)\leq-1$ for all
$t>0$\},

$\beta^{+}~=~\{n\in{R}|~ V_t(t;n)>0$ for all $t\geq0$ and $V(t;n)>
-1$ for some $t>0$ \}.

\begin{lem} We have the disjoint union
$R~=~\beta^{-}\cup\beta^{0}\cup\beta^{+}$.
\end{lem}
\begin{pf} If $n\not\in\beta^{-}$, then
$V_t(t;n)\geq{0}$ for all $t$. If there exists a point $t_0>0$ so
that $V_t(t_0;n)=0$, then $V(t_0;n)\neq~0$ because $V(t;n)=0$ is an
equilibrium point of the differential equation in (4.3) which is not
attainable in finite time. Using the information $V_t(t_0;n)~=~0$
but $V(t_0;n)~\neq~0$ in (4.3), we see that either $V''>0$ or
$V''<0$ at $t~=~t_0$. Hence, there is a $t>t_0$ or $t<t_0$ at which
$V_t(t;n)<0$. This contradicts the assumption that
$n\not\in\beta^{-}$. Thus $V_t(t;n)~>~0$ for all $t>0$ and
$n\in\beta^{0}\cup\beta^{+}$, which proves the relation
$R~=~\beta^{-}\cup\beta^{0}\cup\beta^{+}$ as claimed.
\end{pf}

\begin{lem}
The set $\beta^{+}$ and $\beta^{-}$ are both open and nonempty.
\end{lem}

\begin{pf} The fact that $\beta^{-}\neq \emptyset$ follows immediately from the fact that $(-\infty, 0)\subset\beta^{-}$. To see that $\beta^{+}$ is nonempty, we integrate (4.3) to get
$$
V_t(t;n)~=~(n+\int_0^{t}R(V(s_{1};n)e^{3s_{1}}d s_{1})e^{-3t},
 \eqno(4.4)
$$
$$
V(t;n)~=~m+n(1-e^{-3t})+\int_0^{t}e^{-3s_{1}}(\int_0^{s_{1}}R(V(s_{2};n))e^{3s_{2}}d{s_{2}})d{s_1}.
 \eqno(4.5)
$$
For any fixed $t_{0}>0$, we can choose $n>0$ sufficiently large so
that
$$
V_{t}(t_{0};n)>0,
 \eqno(4.6)
$$
$$
V(t_{0};n)>-1.
 \eqno(4.7)
$$
Considering $V_{0}(0;n)=n>0$, $V(0;n)=m<-1$~and the property of
continuous function, we can see that there exist a set of intervals
$\{(0,\delta_{n})\}$ so that $(0,\delta_{n})\subset(0,\delta_{n+1})$
and $V_{t}(t;n)>0$  for all $t\in(0,\delta_{n})$ where $n\in Z^{+}$.
As the basis for the proof of this lemma, we will first prove that
there exists $k\in Z_{+}$ so that if we denote
$t_{1}=min\{\delta_{k},t_{0}\}$, there holds $V_{t}(t;n)>0$ for all
$t\in(0,t_{1}]$ and $V(t;n)<-1$ for all $t\in(0,t_{1})$ but
$V(t_{1};n)=-1$. Suppose otherwise that there exists no such $t_{1}$
satisfying the condition mentioned closely above. Denote
$T=sup\{\delta_{n}\}$ and it is easy to see that $V_{t}(T;n)=0$,
thereby $T\neq t_{0}$ because $V_{t}(t_{0};n)=0$. Therefore, we can
divide the proof of the lemma into two sections according to whether
$T<t_{0}$ or $T>t_{0}$.

First, if $T<t_{0}$, then the supposition mentioned closely above
leads to $V(T;n)<-1$ and $V_{t}(T;n)=0$. Moreover, from the property
of continuous function, we know that there exists $T_{0}>0$~so that
$V(t;n)<-1$,$\forall t\in (T,T+T_{0})$. Therefore, it is easy to
conclude that
$$
V_{t}(T+\frac{T_{0}}{2})<0,
 \eqno(4.8)
$$
$$
V(T+\frac{T_{0}}{2})<-1.
 \eqno(4.9)
$$
Clearly, the two inequities listed above together with the structure
of $V_{t}(t;n)$ and $V(t;n)$ listed above allow us to see that for
any $t\in(T+\frac{T_{0}}{2},t_{0})$ there hold
$$
V_{t}(T+\frac{T_{0}}{2})<0,
 \eqno(4.10)
$$
$$
V(T+\frac{T_{0}}{2})<-1.
 \eqno(4.11)
$$
which contradict (4.6) and (4.7), thus the lemma is proved provided
that $T<t_{0}$.

Second, if $T> t_{0}$, the supposition mention above allows us to
see that for any $t\in[0,t_{0}]$ there holds
$$
\begin{array}{lr}
V(t;n)<-1,
\end{array}
 \eqno(4.12)
$$
which also contradicts (4.7).

Considering the two cases, we can see that there exists
$\delta_{k}\in (0,t_{0})$(We denote this $\delta_{k}$ as $t_{1}$) so
that $V_{t}(t;n)>0$ for any $t\in[0,t_{1}]$ and $V(t;n)<-1$,
$\forall t\in[0,t_{1})$ but $V(t_{1};n)=-1$.

We will then prove that $V_{t}(t;n)>0$ for all $t\in[0,\infty)$. In
fact, suppose otherwise that there exists $t_{3}>t_{1}$ so that
$V_{t}(t;n)>0$ for any $t\in[t_{1},t_{3})$ but $V_{t}(t_{3};n)=0$.
Noting that $V(t;n)>-1$, $\forall t\in[t_{1},t_{3}]$ and considering
(4.4), we can see that there holds
$$
n+\int_0^{t_{3}}R(V(t;b))e^{3t}dt>n+\int_0^{t_{1}}R(V(t;b))e^{3t}dt>0.
 \eqno(4.13)
$$
Therefore,
$$
(n+\int_{0}^{t_{3}}R(V(\tau;b))e^{3\tau}d \tau )e^{-t_{3}}>0,
 \eqno(4.14)
$$
which contradicts the fact that $V_{t}(t_{3};n)=0$. Hence, we know
that $V(t;n)>0$,$\forall t\in[0,\infty)$. Consequently,we can
conclude that $V(t;n)>-1$ for all $t\in(t_{1},\infty)$, and
naturally, we can see that $V_{t}(t;n)>0$ for all
$t\in(t_{1},\infty)$. Therefore, $n\in\beta^{+}$ and the
nonemptyness of $\beta^{+}$ is established.

Moreover, for $n_{0}\in\beta^{+}$, there is a $t_{0}>0$ so that
$V(t_{0};n_{0})>-1$. By the continous dependence of $V$ on the
parameter $n$ we see that when $n_{1}$ is close to $n_{0}$ we have
$V_{t}(t;n_{1})>0$ for all $t\in[0,t_{0}]$ and $V(t_{0},n_{1})>-1$.
As proved above, $V_{t}(t;n)>0$ for all $t>t_{1}$. Thus, we can see
that $V_{t}(t;n_{1})>0$ for all $t>t_{0}$ as well, which proves
$n_{1}\in\beta^{+}$. So $\beta^{+}$ is open. The fact that
$\beta^{-}$ is open is self-evident. The lemma follows.
\end{pf}
\begin{lem}
The set $\beta^{0}$ is a nonempty closed set. Moreover, if
$n\in\beta^{0}$, then $V(t;n)<-1$ for all  $t>0$.
\end{lem}
\begin{pf}The first part of the lemma follows from the
connectedness of $R$ and Lemma $4.2$. To prove the second part, we
assume otherwise that there is a $t_{0}>0$ so that $V(t_{0};n)=0$.
Since $V(t;n)\leq-1$ for all $t>0$, $V$ attains its local maximum at
$t_{0}$. In particular, $V_{t}(t_{0};n)=0$, which contradicts the
definition of $\beta^{0}$.
\end{pf}
\begin{lem}
For $n\in\beta^{0}$, we have $V(t;n)\to -1$ as $t\to \infty$.
\end{lem}
\begin{pf} Since $V$ increases and $V<-1$ for all $t>0$, we see
that the limit $\lim\limits_{t\to \infty}V(t;n)=V_{\infty}$ exists
and $-\infty<V_{\infty}\leq0$. If $V_{\infty}<0$, then
$R(V(t;n))<R(V_{\infty})<0$. Inserting this result into (4.4), we
see that $V_{t}(t;n)<0$ when $t>0$ is sufficiently large, which
contradicts the definition of $\beta^{0}$.
\end{pf}
\begin{lem}
The set $\beta^{0}$ is actually a single point set. In other words,
the correct shooting data is in fact unique.
\end{lem}
\begin{pf}Suppose otherwise that there are two points
$n_{1}$ and $n_{2}$. Let $V(t;n_{1})$ and $V(t;n_{2})$ be the
corresponding solutions of (4.3). Then the function
$w(t)=V(t;n_{1})-V(t;n_{2})$ satisfies the boundary condition
$w(0)=w(\infty)=0$ and the equation
$$
w''(t)+w'(t)=R'(\xi(t))w(t),~0<t<\infty,
 \eqno(4.15)
$$
where $\xi(t)$ lies between $V(t;n_{1})$ and $V(t;n_{2})$ and
$R'(V)=\frac{d R(V)}{d V}>0$($\forall V$) in view of (3.10) and
(3.11). Applying the maximum principle to (4.15), we conclude that
$w(t)~\equiv~0$, which contradicts the assumption that $n_{1}\neq
n_{2}$.
\end{pf}
For $n\in\beta^{0}$, we now consider the decay rate of $V(t;n)$ as
$t\to \infty$. To simplify our problem, we introduce the following
transformation $v=V+1$. From the properties of the function $R(V)$,
we see that the linearized equation of the differential equation in
(4.3)  around $v=0$ is $\theta''+3\theta'-4\theta=0$, whose
characteristic equation has the roots $\lambda=-4$ and $\lambda=1$.
Hence, we see that for any $\epsilon\in(0,1)$, there is a constant
$C(\epsilon)$ such that
$$
\begin{array}{lr}
-C(\epsilon)e^{-4(1-\epsilon)t}<v(t;n)<0,\forall t\geq0.
\end{array}
 \eqno(4.16)
$$
Note that, modulo the positive small constant $\epsilon$, the above
estimate is sharp. We now go back to the variable $s=-t$. Thus, we
have obtained a solution $V(s)$ of (3.12) defined in the left of the
real line, $-\infty<s\leq0$, such that $V(s)\leq-1$ for all
$s\leq0$, and
$$
-1-C(\epsilon)e^{4(1-\epsilon)s}<V(s)<-1, ~~\forall~s\leq0.
 \eqno(4.17)
$$

We now consider the right half of the real line, $0\leq s<\infty$.
When $s$ is near zero, there hold $V'(s)<0$ and $V(s)<-1$. Inserting
these into (4.1) and using (3.11), we see that $V''(s)<0$ there.
This property implies that the structure of the differential
equation in (3.11) allows us to preserve the negative sign for all
$V(s)$, $V'(s)$ and $V''(s)$. In particular, the solution $V(s)$
exists for all $s>0$ and $V(s)$ is strictly decreasing everywhere.
From the structure of the function $R(V)$ on the right-hand side of
the differential equation, we easily deduce that
$V(\infty)=-\infty$.
 Hence, a solution of (3.12) is obtained. We now strengthen our
conclusion by deriving the accurate blow-up rate for $V(s)$ as $s\to
\infty$.

Integrating the differential equation in (4.1), we obtain
$$
e^{-3s}V'(s)=-n-2\int_{0}^{s}(1-e^{Q(V)})^{2}e^{-3s_{1}}d s_{1}.
\eqno(4.18)
$$
From (3.10) we can see that the integral on the right-hand side of
(4.18) is convergent for $s\to \infty$. Thus, we have the sharp
expression
$$
V'(s)=-(n+\sigma(s))e^{3s}, \eqno(4.19)
$$
where $\sigma(s)=2\int_{0}^{s}(1-e^{Q(V)})^{2}e^{-3s_{1}}d s_{1}$ is
a bounded increasing function in $[0,\infty)$ and $\sigma(0)=0$.
Consequently, we find that $V(s)$ has the following asymptotic
behavior
$$
V(s)=-(n+\sigma(s))e^{3s},~s\geq0. \eqno(4.20)
$$
In other words, the function $V(s)$ blows up to $-\infty$ as fast as
the function $-e^{3s}$ as $s\to \infty$.

We need also get the asymptotic behavior of $V'(s)$ as $s\to
-\infty$. For this purpose, consider the representation (4.3) in
terms of the variable $t=-s$. Using the estimate (4.16) and (3.9),
we see that the factor in front of $e^{-3t}$ on the right-hand side
of (4.4) is bounded. This establishes $V_{t}=O(e^{-3t})$ as $t\to
\infty$. Therefore, we obtain the asymptotic estimate
$$
V'(s)=O(e^{3s}),~s\to -\infty.
 \eqno(4.21)
$$
It is clear that Lemma $5$ implies some kind of uniqueness property
for the boundary value problem (3.12). More precisely, we state

\begin{lem}
Up to translations, $s\mapsto s+s_{0}$, the two-point boundary value
problem (3.12) has a unique solution.
\end{lem}

\begin{pf}
Let $V_{1}$ and $V_{2}$ be two solutions of (3.12). Then they are
all negative-valued and strictly decreasing and their behavior
indicates that there exists a unique point $s_{0}$ so that
$V_{1}(0)=V_{2}(s_{0})$. Set $V_{3}(s)=V_{2}(s+s_{0})$, then both
$V_{1}$ and $V_{3}$ are solutions of the differential equation in
(3.12) and $V_{1}(0)=V_{3}(0)$. Using lemma $4.5$, we have
$V'_{1}(0)=V'_{3}(0)$. Applying the uniqueness theorem for the
initial value problem of an ordinary differential equation, we have
$V_{1}~\equiv~V_{3}$, namely, $V_{1}(s)=V_{2}(s+s_{0})$ for all $s$
and the lemma follows.
\end{pf}
Now consider the boundary behavior of the function $G=Q(V)$. Using
$\frac{d G}{d V}=Q'(V)=\frac{1}{(1-e^{G(V)})}$, $G\to -\infty$ as
$V\to -\infty$ and the L'Hospital rule, we have
$$
\lim\limits_{s\to
\infty}\frac{G(s)}{V(s)}=\lim\limits_{s\to\infty}\frac{1}{(1-e^{Q(V)})}=1.
 \eqno(4.22)
$$
Combining (4.20) and (4.22), we see that for any $\epsilon>0$ there
is a number $S_{\epsilon}>0$ so that
$$
(1+\epsilon)V(s)\leq G(s)\leq(1-\epsilon)V(s),~s\geq S_{\epsilon}.
 \eqno(4.23)
$$
With this estimate, we can consider $G'(s)$ in terms of $V'(s)$ when
$s\to \infty$. Indeed, using the relation between $G(s)$ and $V(s)$,
we have, for sufficiently large $s>0$,
$$
G'(s)=(1-e^{G(s)})^{-1}V'(s)=(1+e^{G(s)}+O(e^{2G(s)}))V'(s).
 \eqno(4.24)
$$
Similarly, we need to consider the asymptotics of $G(s)$ and $G'(s)$
as $s\to -\infty$. Using the relation $V=G-e^{G}$, we have
$$
V=-1-\frac{1}{2}G(s)^{2}+O(G(s)^{3}).
 \eqno(4.25)
$$
for $G(s)$ near zero. Applying (4.17) in (4.25), we obtain the
estimate
$$
-C(\epsilon)e^{2(1-\epsilon)s}<G(s)<0,
 \eqno(4.26)
$$
where $\epsilon>0$ can be made arbitrarily small and $C(\epsilon)>0$
is a constant depending on $\epsilon$. Note again that, modulo
$\epsilon$, the estimate (4.26) is sharp. In terms of $t=-s$,
$G(t)=O(e^{-2(1-\epsilon)t})$ as $t\to \infty$. Inserting this into
(4.3) and noting that
$R(V)=-2(1-e^{Q(V)})^{2}=O(e^{-4(1-\epsilon)t})$, we see that
$V_{tt}+3V_{t}=O(e^{-4(1-\epsilon)t})$. From this we get the
estimate $V(t)=O(e^{-4(1-\epsilon)t})$. Note that (4.3) indicates
that $V_{tt}>0$, since $V_{t}>0$. Hence $V_{t}$ is decreasing.
Therefore, there holds
$$
V_{t}<V(t)-V(t-1)=O(e^{-4(1-\epsilon)t}),~t\geq 1.
 \eqno(4.27)
$$
Consequently, returning to the variable $s=-t$, we obtain the
improved estimate
$$
V'(s)=O(e^{4(1-\epsilon)s}),~as~ s\to -\infty,
 \eqno(4.28)
$$
over (4.21). Inserting this result into the relation
$$
G'(s)=(1-e^{G(s)})^{-1}V'(s)=-(1+O(G(s)))^{-1}(G(s))^{-1}V'(s),
 \eqno(4.29)
$$
we acquire the asymptotic estimate
$$
G'(s)=O(e^{2(1-\epsilon)s}),~as~s\to -\infty,
 \eqno(4.30)
$$
which is compatible with (4.26).

We will then return to the original variable $r$ and give the
asymptotic estimate of $U$ and $K$ in terms of $r$. Note that we
once applied the variable transformation
$$
\tau=ar,~s=\ln \tau,
 \eqno(4.31)
$$
and the function transformation
$$
f(s)=K^{2},~G=\ln f,~V=G-e^{G}.
 \eqno(4.32)
$$
Hence, both $U$ and $K$  in the original boundary problem can be
represented with $G$. Applying (4.23) and (4.24), and with the
understanding that the arbitrarily small constant $\epsilon>0$ is
omitted in the final expression to simplify the notation, we arrive
at
$$
K=O(e^{-Cr^{3}}),~C>0,~as~r\to \infty.
 \eqno(4.33)
$$
In fact, the second equation of (2.12) origins from
$$
\frac{d(U-1)}{d\tau}=\frac{(1-K^{2})^{2}}{(\tau)^{4}}, \eqno(4.34)
$$
where $\tau=ar$, $a=(\frac{2g^{2}}{3} H_{0})_{\frac{1}{3}}>0$.
Therefore, when $r\to \infty$, thereby $r\to \infty$, we can get
$$
\frac{d(U-1)}{d\tau}=O((\tau)^{-4}). \eqno(4.35)
$$
It follows immediately that
$$
U=1+O((\tau)^{-3})=1+O(r^{-3}),~ as~r\to\infty. \eqno (4.36)
$$
Similarly, from (4.29) and (4.30) we can acquire
$$
K~=~1+O(r^{2}),~U~=~O(r),~as~r\to0. \eqno (4.37)
$$
The proof of Theorem 2.1 is now complete.

\begin{lem}
The correct shooting slope, $-n<0$, depends on $m$ continuously and
monotonically so that $n(m_{1})>n(m_{2})>0$ for $m_{1}<m_{2}<-1$.
\end{lem}
\begin{pf}
We have seen that for any given $m<-1$, there is a unique number
$n>0$ so that the unique solution of the initial value problem (4.1)
gives a negative valued solution $V$ which solves the two-point
boundary value problem (3.12)(cf.lemma4.5). Thus we can denote this
well-defined correspondence as $n=n(m)$ and $V=V_{m}$. We show that
$n(m)$ is continuous with respect to $m<-1$. Let $\{m_{j}\}$ be a
sequence in $(-\infty,-1)$ which converges to a number $m_{0}<0$. We
need to prove that $n(m_{j})\to n(m_{0})$ as $j\to \infty$. Suppose
otherwise that this is not true. Then, without loss of generality,
we may assume that there is an $\epsilon_{0}>0$ so that
$|n(m_{j})-n(m_{0})|\geq\epsilon$ for all $j=1,2,...$. On the other
hand, we can use lemma 4.6 to obtain a sequence $\{s_{j}\}$ so that
$V_{m_{j}}(s)=V_{m_{0}}(s_{j}+s)$ for all $s$. In particular,
$m_{j}=V_{m_{j}}(0)=V_{m_{0}}(s_{j})$ for $j=1,2,....$ It is clear
that $\{s_{j}\}$ is a bounded sequence otherwise it would contradict
the assumption $m_{j}\to m_{0}<0~(j\to0)$ and the fact that
$V_{m_{0}}(-\infty)=-1$ and $V_{m_{0}}(\infty)=-\infty$. By
extracting a subsequence if necessary, we may assume that $s_{j}\to
some~s_{0}$ as $j\to \infty$. Therefore, we have, as $j\to \infty$,
$n(m_{j})=V'_{m_{j}}(0)=V'_{m_{0}}(s_{j})\to
V'_{m_{0}}(s_{0})~:=~n_{0}\neq n(m_{0})$. On the other hand,
$m_{j}\to-1$ as $j\to \infty$, and
$m_{j}=V_{m_{j}}(0)=V_{m_{0}}(s_{j})$ for $j=1,2,...$ imply that
$s_{j}\to 0$ as $j\to \infty$ since $V_{m_{0}}$ is strictly
monotone. Hence $s_{0}=0$ and we arrive at a contradiction.

The continuous dependence of $n(m)$ on $m$ implies that the solution
$V_{m}$ depends on $m$ continuously as well.We claim that $n(m)\to
0$ as $m\to 0^{-}$. Otherwise there is a sequence $\{m_{j}\}$ in
$(-\infty,-1)$ and an $\epsilon_{0}$ so that $m_{j}\to 0$ as $j\to
\infty$ but $n(m_{j})\geq\epsilon_{0}(j=1,2,...)$. Using these in
the initial value problem (4.3) with $m=m_{j}$ and $n=n(m_{j})$, we
observe that the solution will assume a positive value for a
slightly positive $t$ when $j$ is sufficiently large, which
contradicts the definition of $n(m_{j})$.

We can also claim that $n(m)\to \infty$ as $m\to -\infty$. Let
$V_{0}$ be a fixed solution of (3.12). Then there is a unique
$s_{m}$ so that $V_{m}(s)=V_{0}(s_{m}+s)$(cf. lemma6). Since
$m=V_{m}(0)=V_{0}(s_{m})$, we conclude that $s_{m}\to \infty$ as
$m\to -\infty$. Consequently, $n(m)=-V'_{m}(0)$ as $m\to -\infty$ as
claimed.
\end{pf}
{\bf Remarks} \quad Our analysis suggests a dynamical shooting
method for constructing the unique solution of the generalized
monopole problem in six-dimension non-Abelian gauge. We have seen
that we may start from the initial value problem (4.1) with an
arbitrary $m<0$. The sets of undesired shooting data, $\beta^{-}$
and $\beta^{+}$ are two open intervals $\beta^{-}=(-\infty,n)$ and
$\beta^{+}=(b,\infty)$. The correct shooting slope, $-n<0$, depends
on $m$ continuously and monotonically so that $n(m_{1})>n(m_{2})>0$
for $m_{1}<m_{2}<-1$.\\

\label{}



\section*{Acknowledgments}
The research of the first author was supported in part by the
Natural Science Fund of Henan Education Office (2007110004) and
(2008A110002).

\end{document}